\documentclass[twocolumn]{aa}
\usepackage{graphicx}
\usepackage{txfonts}
\usepackage{lscape}
\usepackage{rotating}
\usepackage{longtable}
\begin{document}
   \title{An exceptional population of late-type WC stars
in the metal-rich spiral galaxy M~83
\thanks{Based on observations  made with ESO Telescopes at the 
Paranal Observatory under programme ID 69.B-0125}}

\titlerunning{WR population of M~83}

   \author{Paul A. Crowther
          \inst{1}
          \and L. J. Hadfield\inst{1} \and 
          H. Schild\inst{2} \and W. Schmutz\inst{3}}

   \offprints{P. A. Crowther (Paul.Crowther@sheffield.ac.uk)}

   \institute{ Department of Physics and
Astronomy, University of Sheffield, Hicks Building, Hounsfield Rd,
Shefffield, S3 7RH\\
         \and
             Institut f\"ur Astronomie, ETH-Zentrum, CH 8092 Z\"urich, 
                Switzerland
          \and
Physikalisch-Meteorologisches Observatorium, CH-7260 Davos,
           Switzerland}
\date{Received March 2004/Accepted}

   \abstract{We have surveyed the metal-rich spiral galaxy M~83 (NGC~5236) for
its Wolf-Rayet population using VLT-FORS2 narrow-band imaging and follow-up 
spectroscopy. From a total of 280 candidates identified using He\,{\sc ii}
$\lambda$4686 imaging, Multi Object Spectroscopy of 
198 sources was carried out, revealing 132 objects containing 
bona-fide Wolf-Rayet features. From this sample, an exceptional W-R content
of $\sim$1030 is inferred, with  N(WC)/N(WN)$\sim$1.3, continuing the 
trend to larger values at higher metallicity amongst Local Group galaxies. 
More dramatic is the dominance of late-type  WC stars
in M~83 with N(WC8--9)/N(WC4--7)=9 which we attribute to
the sensitivity of the 
classification line  C\,{\sc iii} $\lambda$5696 to mass-loss, providing 
the strength of WC winds scale with metallicity. 
One young massive compact cluster, \#74 in our catalogue, hosts 20\% of 
the entire galactic  population, namely $\sim$180 late WC stars  and 
$\sim$50  late WN stars. 
   }

   \maketitle
%

\section{Introduction}

Wolf-Rayet stars represent excellent diagnostics of recent star formation in 
starbursts. Though few in number, they possess very powerful winds and so contribute 
significantly to the energy released into the ISM in young starbursts
(Crowther \& Dessart 1998), and are
believed to represent the immediate precursors of Type Ib/c Supernovae and
long duration Gamma-Ray Bursts. Individual 
WR stars are readily identified in nearby galaxies via their characteristic
strong, broad emission lines, whilst their collective presence can be seen
in the integrated light of distant starbursts.

Mass-loss rates of hot, massive stars prior to the WR phase are known to
scale with metallicity (Vink et al. 2001), such that the relative number 
of 
WR stars increases with higher metallicity (Maeder \&
Meynet 1994).  In addition, the distribution amongst
nitrogen-rich WN versus carbon-rich WC subclasses is also known to vary
with metallicity  for Local Group galaxies (Massey \& Johnson 1998).
Studies of resolved WR 
populations at metallicities higher than the Solar value have proved
difficult to date because the inner Milky Way is visibly obscured, whilst 
the other metal-rich Local Group member, M31,
lies at an unfavourable inclination.

Towards this goal, M~83 (NGC~5236) is a massive spiral galaxy
that is believed to be particularly metal-rich:
log(O/H)+12=9.0--9.3 as derived from strong-line nebular methods (Bresolin 
\& 
Kennicutt 2002)\footnote{However, see Kennicutt et al. (2003)
regarding the reliability of strong nebular line techniques at high 
metallicity.} and is the 
focus of the present Letter. 
A more detailed discussion of the WR population of M~83 and presentation
of the full catalogue is given by Hadfield et al. (in preparation).


\section{Observations}\label{sect2}

\subsection{Imaging}

We observed M~83 with the ESO Very Large
Telescope UT4 and Focal Reducer/Low
Dispersion Spectrograph \#2 (FORS2) between Apr--Jun 2002.
FORS2 has a 6.8\,\arcmin$\times$6.8\,\arcmin\ field-of-view using the
standard collimator, with an image scale of 0.2\,\arcsec/pixel.
FORS2 was used to obtain narrow-band $\lambda$4684 (C\,{\sc iii}/He\,{\sc 
ii}, FWHM=66\AA) and $\lambda$4781 (continuum, FWHM=68\AA) images of M~83 
in order
to detect WR candidates which possess net $\lambda$4686 emission,
with exposure times of 1800~s.

Since M~83 subtends 12.9$'$ by 11.5$'$ on the sky, the entire
galaxy could not be imaged using a single FORS2 frame. Consequently,
four overlapping fields were selected in order to cover the entire galaxy, 
each $\sim$210 arcsec 
from the nucleus.
WR line and adjacent continuum images for
each field were taken concurrently, with seeing conditions better
than 0.8$''$.

In addition to our  He\,{\sc ii} narrow-band imaging, 
we also acquired broad-band Bessell B (120 and 600s) plus
on- and off- narrow-band H$\alpha$ (60 and 600s) 
images. The data were de-biased, cosmic-ray cleaned and 
flat-fielded, and calibrated 
using images of either a standard field (Bessell B filter) or 
via images of a suitable spectrophotometric standard (narrow-band filters).
Photometry was carried out using {\sc daophot}. We estimate a  photometric 
accuracy of $\sim$0.1 mag for most sources from comparison between 
identical sources observed within different fields.

We obtained difference images from 
our $\lambda$4684 and $\lambda$4781 frames, from which a large number of
candidate $\lambda$4684 emission sources were identified, following
the technique of Schild et al. (2003). In total 280 
candidates were found, for which photometry in (at least) the  $\lambda$4684 
filter was obtained in 75\% of cases. Severe crowding was a major
limitation with obtaining reliable photometry for the remainder of cases.

\subsection{Spectroscopy}

We used FORS2 in Multi Object Spectroscopy (MOS) mode 
during Apr--June 2003 to obtain spectroscopy
of individual M~83 candidates using the 300V grism, 
centred at $\lambda$=5900\AA. Up to 19 objects may in principal be 
observed
simultaneously, whilst in practice typically 15 candidates could be suitably
positioned. A total of 198 candidates were observed 
spectroscopically
with FORS2 in 17 settings, each with a 0.8 arcsec slit, during 
seeing conditions superior to 0.8$''$. 
The 2 pixel spectral resolution
obtained was $\sim$7\AA. On-source exposures ranged from 
720~s for the brightest sources to 4800~s for the faintest,
each split into 2--3 individual exposures.

A standard reduction technique was applied using {\sc iraf}. For most
sources an absolute flux calibration was achieved by comparing the 
photometry of an individual object
in the  $\lambda$4684 filter with a synthetic photometric measurement, 
determined by convolving our spectroscopy with a suitable Gaussian filter
and zero-point. The average slit loss factor was 3.1 from 160
sources brighter than 24.0 mag. This factor was adopted for several
spectroscopically observed sources for which photometry was unavailable.

 \section{Analysis}\label{sect3}

We have inspected our 198 candidates for which spectroscopy was obtained.
In 132 cases we have identified genuine, broad WR emission features, 
namely
either the blue $\lambda$4647-51 C\,{\sc iii} -- $\lambda$4686 He\,{\sc ii}
blend, and/or yellow $\lambda$5696 C\,{\sc iii} or $\lambda$5801-12 
C\,{\sc iv} lines.  For the remaining 66 cases in which spectroscopy
was obtained, either no emission was identified or the object was a
foreground, late-type star. 
For a small number of cases either the S/N achieved from spectroscopy was 
insufficient to 
confirm the presence of WR features, or the dataset started longward
of the blue visible region, necessary for identification of WN stars. 

\begin{table}[ht]
\caption[]{Total distribution of WR stars in M~83
inferred from de-reddened line
fluxes  based on a distance of 4.5~Mpc (Thim et al, 2003) and
line luminosity (in erg s$^{-1}$) calibration adapted from Schaerer \& 
Vacca (1998). The total number of regions add up to greater than 132 since 
5 regions contain both WC and WN stars.
\label{table1}}
\begin{small}
\begin{tabular}{lrrrrr}
\noalign{\hrule\smallskip}
Subtype & WNE & WNL & WC4--6 & WC7 & WC8--9 \\
Line & $\lambda$4686 & $\lambda$4686 & 
$\lambda$5801-12 & $\lambda$5801-12 & $\lambda$5696 \\
$\log L$(Line) & 35.72 & 36.20 & 36.20 & 36.15 & 35.85 \\
\noalign{\hrule\smallskip}
Regions &  20 & 32  & 13 &  8 &  64 \\
N(WR)   & 220 & 229 & 29 & 26 & 525 \\
\noalign{\hrule\smallskip}
\end{tabular}
\end{small}
\end{table}
Our aim was to characterize the WR population in M~83. Consequently, we 
have attempted to measure intrinsic WR line fluxes and estimate the total 
number of WR stars in each object from comparison with Galactic 
counterparts. A distance to M~83 of 4.5~Mpc was adopted from Cepheid 
measurements of  Thim et al. (2003).
For 70\% of our 132 sources for which WR features were identified,
interstellar reddenings were derived from nebular H$\alpha$ and H$\beta$
lines measured from the extracted spectra. Assuming Case B recombination
theory for typical electron densities 10$^{2}$ cm$^{-3}$ and temperatures
10$^{4}$ K (Hummer \& Storey 1987), we obtain 
0.1 $\leq$ E(B-V)=$c$(H$\beta$)/1.46 $\leq$ 1.0~mag.
Estimated accuracies are typically 
$\pm$0.1 mag, although reddenings for regions with weak 
H$\beta$ will likely represent overestimates given the  neglect of stellar 
H$\beta$ absorption. In 31 cases no nebular lines were observed, in 
which case an estimate of E(B-V) 
could be made from a comparison with a theoretical energy distribution for 
a late O star (Kurucz 1992). In 11 cases with neither nebular lines 
nor a  clear continuum, our average reddening of E(B-V)=0.43 was adopted.

The WR content of individual objects are grouped into either early or
and late subtypes as follows.
WNL subtypes are identified if He\,{\sc ii} $\lambda$4686 is accompanied
by N\,{\sc iii}  $\lambda$4634--41, and WNE if He\,{\sc ii} $\lambda$4686
is accompanied with N\,{\sc v} 
$\lambda$4603--20 or FWHM(He\,{\sc ii}) $\geq$20\AA. 
Amongst WC subtypes, we 
have indicated WC4--6 if C\,{\sc iv} $\lambda$5801--12 is present, with
C\,{\sc iii} $\lambda$5696 weak or absent, WC7 if 
0.25 $\leq$ $F_{\lambda}$(C\,{\sc iii})/$F_{\lambda}$(C\,{\sc iv}) $\leq$ 
0.8, and WC8--9 otherwise.
Total numbers of WR stars in each object were determined using
the calibration of Schaerer \& Vacca (1998).
It is of course possible 
that our  choice of representative line  luminosities
is inappropriate to the environment of M~83, however many Galactic WR 
stars, particularly late WC stars are located towards the Galactic Centre
(see discussion in Schaerer et al. 2000). 

   \begin{figure}
   \begin{center}
   \includegraphics[bb=30 190 525 740,width=\columnwidth,clip]{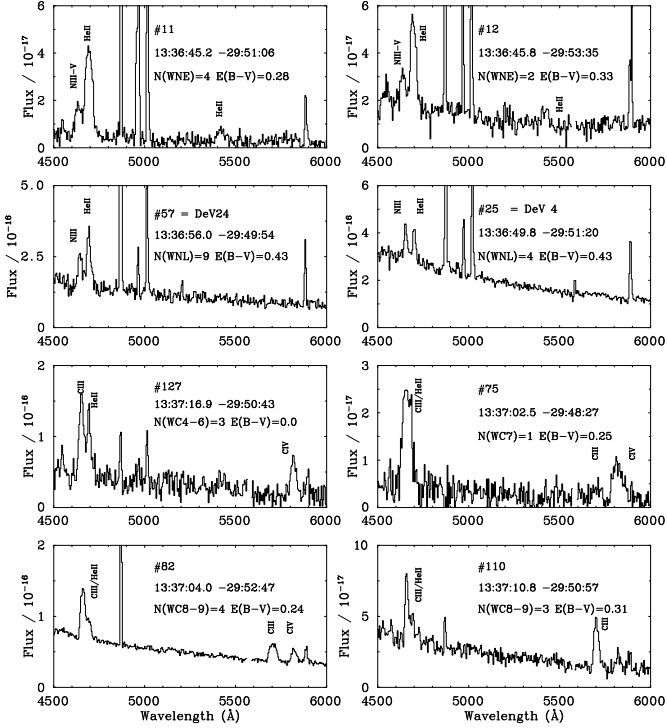}
      \caption{Representative examples of sources containing early WN, 
late WN, early WC and late WC stars in M~83 including J2000 coordinates,
and association with clusters from De Vaucouleurs et al. (1983, DeV)
where available.}
         \label{fig1}
         \end{center}
   \end{figure}

\section{WR population of M~83}\label{sect4}

Representative WR spectra are presented 
in Fig.~\ref{fig1}. In some cases a single WR star is 
identified, whilst others typically
contain $\sim$10, numbers comparable to WR populations in
nearby giant H\,{\sc ii} regions (Drissen et al. 1993).
Table~\ref{table1} summarizes the derived population of WR stars
in M~83 obtained from our analysis. We identify a total of 1030 WR
stars from our 132 spectroscopically confirmed regions, ranging from
$M_{\rm B}=-6$ to $-$14 mag, with $M_{\rm B}=-9$ 
mag on average. Our statistics rely solely upon those sources for which 
spectroscopic
follow-up was obtained. Since we have attempted to observe a genuinely 
representative sample, we expect $\sim$50 our of the remaining 82
candidates to contain genuine WR features, 
such that the total WR population of M~83 may approach 1500. Indeed, one
of our remaining candidates is M83-5 from Bresolin \& Kennicutt (2002) 
who confirmed WN and WC features via spectroscopy.

\subsection{The WN/WC ratio and late WC population}

Massey (1996) first highlighted the sensitivity of N(WC)/N(WN) to metallicity.
As metallicity increases, the relative fraction of WC stars increases from
$\leq$0.1 in the SMC  
to $\sim$0.7 in the Milky Way (van der Hucht 2001). 
For M~83, we obtain N(WC)/N(WN)$\sim$1.3 which
fits in rather well with the extrapolation of previous determinations.
WC stars are, of course, more readily identified in external galaxies 
than WN stars due to their intrinsically stronger lines. Nevertheless, our 
approach is optimised for net emission at $\lambda$4686, such that we
have 4$\sigma$ spectroscopic WNL detections with $W_{\lambda}$(He\,{\sc 
ii} $\lambda$4686) $\sim$ 1\AA.

The most dramatic finding from Table~\ref{table1} is that half of the
total WR population in M~83 are of WC8--9 subtype. To put this into 
context, 18\% of the known WR population comprise WC8--9 stars (van der 
Hucht 2001),
whilst no WC late stars are known elsewhere except a handful in M31 (Moffat
\& Shara 1987). We obtain a remarkably small 
population of WCE stars
in M~83, such that N(WC8--9)/N(WC4--7)$\sim$9, versus 
0.9 in the  Milky Way, $\sim$0.2 in M31 (Massey \& Johnson 1998) 
and 0.0  elsewhere. We  confirm previous Local Group studies of WR
stars that WCL stars are uniquely located in metal-rich environments. 
In our conclusions we shall attempt to explain this subtype behaviour. 
In contrast, the WN population of M~83 is rather more familiar. The
relative fraction of late WN stars to early WN stars is rather low in 
metal poor galaxies such as the SMC and high in the Milky Way, with 
N(WNL)/N(WNE)$\sim$1.6 (van der Hucht 2001). We find 
a very similar ratio of N(WNL)/N(WNE)=1.3 in M~83. 

\subsection{Stellar clusters with large numbers of WR stars}

\begin{figure}
   \centering
   \includegraphics[bb=60 210 490 745,width=0.8\columnwidth,clip]{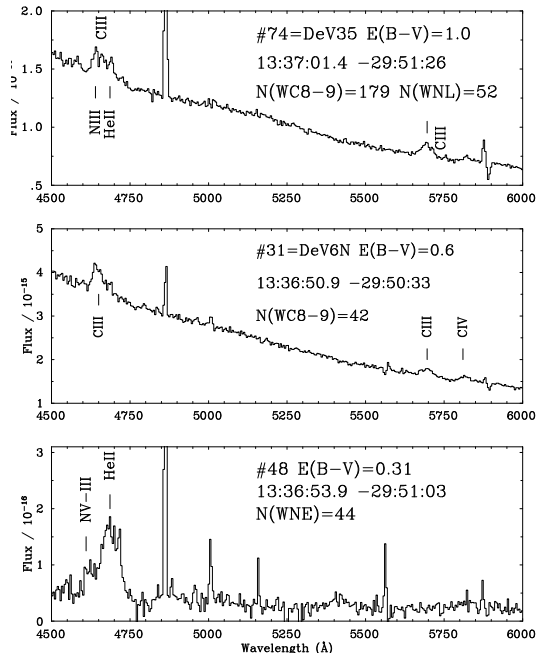}
      \caption{De-reddened, radial velocity corrected (516 km\,s$^{-1}$), 
FORS2 spectroscopy of three clusters in M~83 containing many WR 
stars.}
         \label{fig2}
   \end{figure}

Three regions contain large ($\geq$40) numbers of WR stars. Properties of
these clusters are listed in Table~\ref{table2} with de-reddened
FORS2 spectroscopy presented in Fig.~\ref{fig2}. 
The most exceptional of these is \#74 from our catalogue 
(Hadfield et al.  in 
preparation).  This source 
hosts up to 179
 WCL stars 
from C\,{\sc iii} $\lambda$5696\footnote{A reduced number of 82 
WCL stars
is inferred from C\,{\sc iii} $\lambda$4650, assuming a line
luminosity of 1$\times 10^{36}$ erg\,s$^{-1}$ (Schaerer \& Vacca 1998).}
 plus 52 WNL stars from He\,{\sc ii} $\lambda$4686.
Bresolin \& Kennicutt (2002) failed to detect
WR stars in this region (their region \#8) since the WR excess is
offset by several arcsec N-E of the peak H$\alpha$ emission.  
We have derived a net integrated, de-reddened, 
H$\alpha$ flux of  4.6$\times 10^{-12}$ erg s$^{-1}$ cm$^{-2}$, equal
to $\sim$810 O7V star equivalents (Vacca 1994) for the cluster as a 
whole. From comparison with evolutionary synthesis models for a 5 Myr 
instantaneous burst (Leitherer et al. 1999) we estimate
a mass of 2$\times 10^{5} M_{\odot}$ for \#74. Inspection of archival 
HST/Advanced Camera for Surveys (ACS) Wide Field Camera (WFC) datasets 
of M~83 reveals \#74 to be very  compact, with a diameter 
of $\leq$0.2 arcsec or 5\,pc at the distance of  M83, such that it is a 
young massive compact cluster (or Super Star Cluster, Whitmore 2003) 
instead of a conventional giant H\,{\sc ii} region.

\begin{table}[ht]
\caption[]{M~83 clusters containing large (N(WR)$\geq$40)
numbers of WR stars. Observed line fluxes are expressed in units of erg 
s$^{-1}$cm$^{-2}$
and are derived from our spectrophotometry with the exception of H$\alpha$
which originates from our continuum subtracted images.
\label{table2}}
\begin{small}
\begin{tabular}{lrrrr}
\noalign{\hrule\smallskip}
Catalogue \#   &   74        & 31   & 48\\
\noalign{\hrule\smallskip}
B (mag)&  18.1           &18.3 & 21.0\\
E(B-V) &  1.0            & 0.6 & 0.31\\
$M_{\rm B}$ & --14.2      & --12.4 & --8.6\\
F(N\,{\sc iii-v} 4603-40)  & 1.1$\times 10^{-15}$ & --                  & 
4.5$\times 10^{-16}$\\
F(C\,{\sc iii} 4647-51)  & 1.0$\times 10^{-15}$ &2.4$\times 
10^{-15}$ &--\\
F(He\,{\sc ii} 4686)  & 1.0$\times 10^{-15}$ &8.4$\times 10^{-16}$ 
&3.2$\times 10^{-15}$  \\ 
F(C\,{\sc iii} 5696)  & 3.5$\times 10^{-15}$ &2.4$\times 10^{-15}$ 
&--\\
F(C\,{\sc iv} 5801-12)  & 1.4$\times 10^{-15}$ & 7.1$\times 10^{-16}$ 
&--\\
F(H$\alpha$) & 4.7$\times 10^{-13}$  &  9.8$\times 10^{-14}$ &7.1$\times 10^{-14}$ \\
N(O7V)   & 810           &70   &  25  \\
N(WNE)   & --            & --& 44\\
N(WNL)   & 52            &   & --\\
N(WC8--9)   & 179        &42 & --\\
\noalign{\hrule\smallskip}
\end{tabular}
\end{small}
\end{table}

Prior to the present study, solely Rosa \& D'Odorico (1986) and 
Bresolin \& Kennicutt (2002) have identified
the presence of WR stars in selected H\,{\sc ii} regions in M~83.
For the 5 regions for which the presence of WR stars was
identified by Bresolin \& Kennicutt, one was in the nucleus for which our
datasets are saturated, the remainder hosted between 1 and 10 late WN stars.
Differences for sources in common are primarily due to the assumption of a smaller 
distance of 3.2\,Mpc by Bresolin \& Kennicutt (2002), plus generally higher
reddenings in the present study.

\section{Summary and Conclusions}\label{sect5}

We have surveyed the nearby metal-rich galaxy M~83 for the presence of 
WR stars using VLT-FORS2. From follow up spectroscopy of 198 out of 280
candidates, 132 regions containing WR stars are identified.
Assuming intrinsic line fluxes comparable to Galactic counterparts (Schaerer
\& Vacca 1998), we identify 1030 WR stars in M~83, with 
N(WC)/N(WN)$\sim$1.3,
continuing the trend observed amongst Local Group galaxies to higher
metallicity (Massey \& Johnson 1998). Accounting for the remaining candidates
the total WR population may be as high as 1500. More than 50\% of the known
WR are identified as late WC stars, versus $\leq$5\% early WC stars, which is
unprecedented relative to more metal-poor Local Group galaxies. One young 
massive compact cluster,
\#74, hosts $\sim$230 late WN and WC stars. 

The relatively large WC population
with increasing metallicity can readily be understood from 
comparison with evolutionary models (Maeder \& Meynet 1994). At 
higher metallicity, mass-loss rates during and subsequent to the main-sequence
evolution of massive stars strips away higher layers earlier on, such that
a star with a particular initial mass advance to later (WC) phases at higher
metallicity. 

But why are there exclusively {\it early } WC stars in the LMC, 
a mixed population in the Milky Way and almost exclusively {\it late} WC stars in
M~83? A decade ago it was believed that (C+O)/He increases from WC9 to WC4
(Smith \& Maeder 1991). However, subsequent spectral analysis failed
to confirm any systematic trend in C/He with subtype 
(Koesterke \& Hamann 1995; Crowther et al. 2002). If carbon abundances do 
not play a dominant role, what does? We suggest differences in wind 
densities are primarily responsible. The wind densities of WC stars in the 
LMC  are  $\sim$50\% lower than those of their  Milky Way counterparts 
according
to Crowther et al. (2002), which they attributed to a metallicity dependence
of their winds. Since C\,{\sc iii} 
$\lambda$5696 increases dramatically in strength with increasing mass-loss 
rate (see their Fig.12), one would expect yet higher mass-loss rates and
even stronger C\,{\sc iii} $\lambda$5696 in M~83 WC stars if WC winds are 
metallicity dependent.
Our present results are fully 
consistent
with such a population. A genuine metallicity dependence of WR winds has 
implications
for the hard ionizing fluxes of young starbursts (Smith et al. 2002).
Regardless,
the presence of late WC stars is definitely a indicator of a metal-rich 
environment. 

Finally,  should  a supernova be observed in M~83 in the near 
future, we now possess  a reasonable  statistical sample with which we 
should be able to verify whether a WR star was a potential immediate 
precursor. 

\begin{acknowledgements}
PAC acknowledges financial support from the Royal Society, LH
acknowledges financial support from PPARC and a Royal Society summer
studentship award.
\end{acknowledgements}


\end{document}